\documentclass[preprintnumbers,amsmath,amssymb]{revtex4}
\usepackage{epsfig}
\usepackage{bm}

\newcommand{\s}{\sigma}
\newcommand{\la}{\lambda}

\newcommand{\vp}{\vec{\pi}}

\begin{document}

\title{Evolution of Critical Correlations at the QCD Phase Transition}

\author{N.~G.~Antoniou}
\author{F.~K.~Diakonos}
\author{E.~N.~Saridakis}
\email{msaridak@phys.uoa.gr} \affiliation{Department of Physics,
University of Athens, GR-15771 Athens, Greece}

\date{\today}

\begin{abstract}

We investigate the evolution of the density-density correlations
in the isoscalar critical condensate formed at the QCD critical
point. The initial equilibrium state of the system is
characterized by a fractal measure determining the distribution of
isoscalar particles (sigmas) in configuration space.
Non-equilibrium dynamics is induced through a sudden symmetry
breaking leading gradually to the deformation of the initial
fractal geometry. After constructing an ensemble of configurations
describing the initial state of the isoscalar field we solve the
equations of motion and show that remnants of the critical state
and the associated fractal geometry survive for time scales larger
than the time needed for the mass of the isoscalar particles to
reach the two-pion threshold. This result is more transparent in
an event-by-event analysis of the phenomenon. Thus, we conclude
that the initial fractal properties can eventually be transferred
to the observable pion-sector through the decay of the sigmas even
in the case of a quench.

\end{abstract}

\maketitle

\section{Introduction}

Experiments of a new generation, with relativistic nuclei, at RHIC
and SPS are currently under consideration with the aim to
intensify the search for the existence and location of the QCD
critical point in the phase diagram of strongly interacting matter
\cite{RIKEN,Antoncern}. Important developments in lattice QCD
\cite{karsch} and studies of hadronic matter at high temperatures
\cite{Antoniou2003} suggest that the QCD critical endpoint is
likely to be located within reach at SPS energies. It is therefore
desirable to explore the range of baryon number chemical potential
$\mu_B=$100-500 MeV by studying collisions at relatively low
energies, $\sqrt{s}\simeq\,$5-20 GeV per nucleon pair. A decisive
observation, in these experiments, associated with the development
of a second-order phase transition at the critical endpoint of QCD
matter is the establishment of power laws in momentum space (self
similarity) in close analogy to the phenomenon of critical
opalescence in QED matter \cite{lesne}. These power laws reflect
the fractal geometry of real space, and a characteristic index of
the critical behavior is the fractal mass dimension $D_f$ which
measures the strength of the order-parameter fluctuations within
the universality class of critical QCD \cite{Antoniou2001}.

The physics underlying the endpoint singularity in the QCD phase
diagram is associated with the phenomenon of chiral phase
transition, a fundamental property of strong interactions in the
limit of zero quark masses. In this case and for a given chemical
potential $\mu_B$ there exists a critical temperature $T_{cr}$
above which chiral symmetry is restored and as temperature
decreases below $T_{cr}$, the system enters into the chirally
broken phase of observable hadrons. It is believed that, for two
flavors and zero quark masses, there is a first-order phase
transition line on the $(T,\mu_B)$ surface at large $\mu_B$
\cite{RW1}. This line ends at a tricritical point beyond which the
phase transitions become of second order. In the case of real QCD
with non-zero quark masses, chiral symmetry is broken explicitly
and the first-order line ends at a critical point beyond which the
second-order transitions are replaced by analytical crossovers
\cite{bergraja99}.

The order parameter of the chiral phase transition is the chiral
field $\Phi=(\s,\vp)$ formed by the scalar, isoscalar field $\s$
together with the pseudoscalar, isovector field
$\vp=(\pi^+,\pi^0,\pi^-)$. Both fields are massless at the
tricritical point and when the symmetry is restored at high
temperatures, their expectation value vanishes,
$\langle\vec{\pi}\rangle=\langle\s\rangle=0$. However, in the
presence of an explicit symmetry breaking mechanism (non-zero
quark masses) the pion and sigma fields are disentangled at the
level of the order parameter of the QCD critical point, which is
now formed by the sigma field alone. The expectation value of the
$\s$-field remains small but not zero near the critical point so
that the chiral symmetry is never completely restored. The
valuable observables in this case are associated with the
fluctuations of the $\s$-field, $(\delta\s)^2\simeq
\langle\s^2\rangle$, and they incorporate, in principle, the
singular behavior of baryon-number susceptibility and in
particular the power-law behavior of the $\s$-field correlator
\cite{hatta03}.

The aim of this work is to study the evolution of critical
correlations during the development of the collision, in the
neighborhood of the QCD critical point. In the initial state we
assume that the system has reached the critical point in thermal
equilibrium so the $\s$-field fluctuations are described by the
$3-D$ Ising critical exponent $\delta$ and in particular by the
fractal dimension $D_f=\frac{3\delta}{\delta+1}$
($\delta\simeq5$). The crucial question from the observational
point of view is whether in the freeze-out regime, which follows
the equilibration stage, the relaxation time-scale $\tau_{rel}$ of
these fluctuations is long enough compared to the time-scale
$\tau_{th}$ associated with the development of a massive
$\s$-field beyond the two-pion threshold ($m_\s\geq 2m_\pi$). Both
time-scales ($\tau_{rel},\tau_{th}$) are characteristic parameters
of the out-of-equilibrium phenomena ($\s$ rescattering) which take
place during the evolution of the system (towards freeze-out) and
the requirement $\tau_{rel}\gg\tau_{th}$ guarantees that critical
fluctuations may become observable in the $\s$-mode
($\s\rightarrow\pi^{+}\pi^{-}$ \cite{Antoniou2001,Fuj}).

In order to quantify these effects, we adopt in this work the
picture of a rapid expansion (quench) which is a realistic
possibility in the framework of heavy-ion collisions. We study the
out-of-equilibrium evolution of the initial fractal
characteristics of the $\s$-field and we search for time scales
$\tau_{rel},\tau_{th}$ satisfying the above constraint in a
particular class of events (event-by-event search). The dynamics
of the system is fixed by a two-field Lagrangian,
$\mathcal{L}(\s,\vp)$, together with appropriate initial
conditions. The out-of-equilibrium phenomena are generated by the
exchange of energy between the $\s$-field and the environment
which consists of massive pions initially in thermal equilibrium.

In section II the formulation of the problem and in particular the
equations of motion, the initial conditions and the generation of
thermal $\pi$-configurations are presented. In section III
numerical solutions of the evolution of critical fluctuations are
given and discussed whereas in section IV our final results and
conclusions are summarized.

\section{Formulation of $\s$ field dynamics}

In our approach we assume an initial critical state of the system
in thermal equilibrium, disturbed by a two-field potential
$V(\s,\vp)$, in an effective description inspired by the chiral
theory of strong interactions \cite{GL,RW}.
 The 3-dimensional Lagrangian
density is
\begin{equation}
\mathcal{L}=\frac{1}{2}(\partial_\mu\s\partial^\mu\s
+\partial_\mu\vp\partial^\mu\vp)-V(\s,\vp) \label{lagr}
\end{equation}
with the potential
\begin{equation}
V(\s,\vp)=\frac{\la^2}{4}(\s^2+\vp^2-v_0^2)^2
+\frac{m_{\pi}^2}{2}\left(\vp^2-2v_0\s+2v_0^2\right), \label{pot}
\end{equation}
where $\s=\s(\vec{x},t)$ and $\vp=\vp(\vec{x},t)$. The potential
has the usual $\s$-model form plus a term which breaks the
symmetry along the $\s$-direction. With the addition of the mass
term for the pion field, we ensure that it has a constant mass
equal to $m_\pi$. Finally, the constant terms in (\ref{pot}) shift
to zero the value of the potential at the minimum. We fix the
parameters of the Lagrangian using the phenomenological values
$m_\pi$$\approx139\,$ MeV and $v_0\approx87.4\,$ MeV, whereas the
known uncertainty in the phenomenological value of zero
temperature $\s$-mass, given by $m_\s=\sqrt{2\lambda^2 v_0^2}$,
yields $10\lesssim\la^2\lesssim 20$ for $400\lesssim m_\s\lesssim
600\,$ MeV.

The equations of motion resulting from (\ref{lagr}) are:
\begin{eqnarray}
\ddot{\s}-\nabla^2\s+\la^2(\s^2+\vp^2-v_0^2)\s=v_0m_\pi^2\nonumber\\
\ddot{\vp}-\nabla^2\vp+\la^2(\s^2+\vp^2-v_0^2)\vp+m_\pi^2\vp=0\label{eom1},
\end{eqnarray}
where $\vp^2=(\pi^+)^2+(\pi^0)^2+(\pi^-)^2$.

Using a constant value $v_0$ in eq. \ref{pot} implies a
non-vanishing mass (finite correlation length) for the $\s$-field,
$m_\s=\sqrt{2\lambda^2 v_0^2}$, in contradiction with its critical
profile. This inconsistency is restored if we assume a finite-time
mechanism instead of the instant quench i.e the instantaneous
formation of the potential (\ref{pot}) at $t=0$. The simplest
model which leaves the equations of motion unaffected, used also
in cosmological phase transitions \cite{Rivers}, is the so-called
linear quench \cite{SRS2}. It assumes that the minimum $v$ of the
potential increases linearly with time, starting from zero and
ending at the zero temperature value $v_0\approx87.4\,$ MeV after
a time interval $\tau$ which is the quench duration:
\begin{equation}
v(t)=v_0 t/\tau  \text{\ \ \ \ \ \ \ \ \ \ for \ \ \ \ \
$t\leq\tau$}\nonumber
\end{equation}
\begin{equation}
 v(t)=v_0   \text{\ \ \ \ \ \ \ \ \ \ \ \ \ \ for \ \ \ \ \ $t>\tau$}.
\end{equation}
In this way we acquire $m_\s=0$ at $t=0$, as expected for the
critical $\s$-field and, with increasing time, $m_\s$ approaches
its zero temperature value.

To proceed numerically we have to discretize eqs.~(\ref{eom1}) on
a lattice. We use the following leap-frog discretization scheme:
\begin{eqnarray}
&\s^{n+2}_{i,j,k}=2\s^{n+1}_{i,j,k}-\s^{n}_{i,j,k}+
\frac{dt^2}{\alpha^2}
\left\{\left[\s^{n+1}_{i+1,j,k}+\s^{n+1}_{i-1,j,k}-2\s^{n+1}_{i,j,k}\right]+
\left[\s^{n+1}_{i,j+1,k}+\s^{n+1}_{i,j-1,k}-2\s^{n+1}_{i,j,k}\right]+\ \ \ \ \ \ \ \ \right.\nonumber\\
&\left.\ \ \ \ \ \ \ \ \ \ \ \ \ \ \ \ \ \ \ \ \ \ \ \ \
\left[\s^{n+1}_{i,j,k+1}+\s^{n+1}_{i,j,k-1}-2\s^{n+1}_{i,j,k}\right]\right\}-
dt^2\left\{\lambda^2\left[\left(\s^{n+1}_{i,j,k}\right)^2+\left(\vp^{n+1}_{i,j,k}\right)^2-v^2\right]
\s^{n+1}_{i,j,k}-vm_\pi^2\right\}, \label{leap-frog}
\end{eqnarray}
and similarly for the other three equations considering the
$\pi$-field. In eq.~(\ref{leap-frog}) $\alpha$ is the lattice
spacing while $dt$ is the time step. The upper indices indicate
the time instants  and the lower indices the lattice sites. As
usual we perform an initial fourth order Runge-Kutta step to make
our algorithm self-starting, and we impose periodic boundary
conditions.

We are interested in studying the evolution of the above system
using initial field configurations dictated by the onset of the
critical behavior. In this case we expect that the $\s$-field,
being the order parameter, will possess critical fluctuations, and
the $\pi$-fields to be thermal, while the entire system will be in
thermodynamical and chemical equilibrium. Obviously, the
subsequent evolution, determined by eqs.~(\ref{eom1}), will
generate strong deviations from equilibrium. Before going on with
the detailed study of the dynamics, we first describe in the
following subsections the generation of an ensemble of field
configurations on a 3-D lattice possessing the characteristics of
the critical system. This ensemble enters in the subsequent
analysis as the initial condition.

\subsection{Generation of initial ensemble of critical $\sigma$-configurations}

The absolute value of the $\s$-field introduced in the previous
subsection is interpreted as local density, and the corresponding
critical behavior is described by a fractal measure demonstrated
in the dependence of the mean "mass" $M(\vec{x}_0,R)$ on the
distance $R$ around a point $\vec{x}_0$ defined by:
\begin{equation}
M(\vec{x}_0,R)=\langle\int_{R}
|\sigma(\vec{x})\sigma(\vec{x}_0)|\, d^Dx\rangle,
\label{massdimen}
\end{equation}
obeying the power law
\begin{equation}
M(\vec{x}_0,R)\sim R^{D_f} \label{masspower}
\end{equation}
for every $\vec{x}_0$. $D_f$ is the fractal mass dimension of the
system \cite{Mandel83,Vicsek,Falconer} and the mean value is taken
with respect to the ensemble of the initial $\s$-configurations.
The production of the $\s$ configurations building up the critical
ensemble, characterized by the fractal measure given in
eqs.~(\ref{massdimen},\ref{masspower}), has been accomplished in
\cite{Antoniou98}. In fact the fractal properties of the critical
system can be produced as an ensemble average, through the
partition function:
\begin{equation}
Z=\int {\cal{\delta}}[\sigma] e^{-\Gamma[\sigma]},\label{partfunc}
\end{equation}
with $\Gamma[\sigma]$ the scale invariant effective action at
$T=T_{cr}$, $\mu=\mu_{cr}$:
\begin{equation}
\Gamma[\sigma]=\int_R d^Dx \{ \frac{1}{2} (\nabla \sigma)^2 + g
\sigma^{\delta+1}  \}. \label{effact}
\end{equation}
The sum over field configurations in (\ref{partfunc}) can be
saturated through saddle point solutions of $\Gamma[\sigma]$,
which consist approximately of piecewise constant configurations
extended over domains of variable size. A detailed description of
the corresponding simulation algorithm is given in
 \cite{manoscomput}. Using this algorithm, after averaging, the
mean "mass" in the ensemble satisfies eq.~(\ref{masspower}) with a
fractal mass dimension
\begin{equation}
D_f=\frac{D\delta}{\delta+1} \label{fracdim}.
 \end{equation}
For the 3-D Ising universality class, $D=3$, the isothermal
critical exponent is $\delta\approx5$, and the coupling
$g\approx2$ \cite{Tsypin}, therefore $D_f\approx5/2$.

\begin{figure}[h]
\begin{center}
\mbox{\epsfig{figure=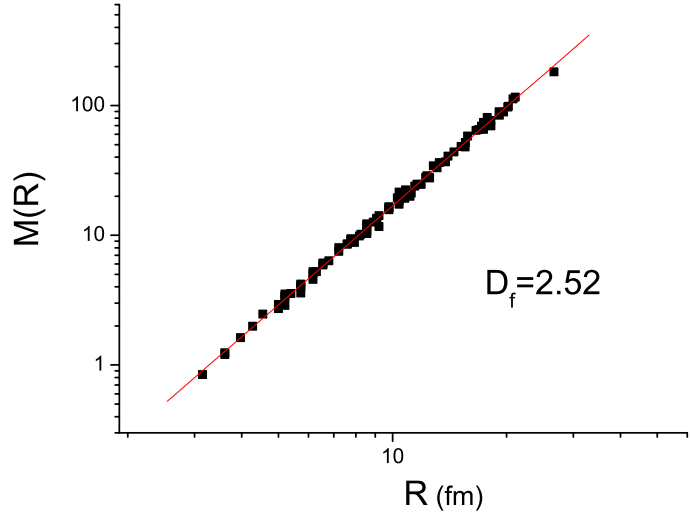,width=10cm,angle=0}} \caption{\it
$M(R)=\langle\int_R |\sigma(\vec{x})\sigma(\vec{x}_0)|\,
d^3x\rangle$ versus $R$ for the ensemble of $\s$-field
configurations. The slope $\Psi$, i.e the fractal mass dimension
$D_f$ is equal to $5/2$  within an error of less than 1\%.}
\label{spower0}
 \end{center}
 \end{figure}

The power-law behavior of $M(\vec{x}_0,R)=\langle\int_R
|\sigma(\vec{x})\sigma(\vec{x}_0)|\, d^3x\rangle$ around a random
$x_0$, averaged inside clusters of volume $V$, is illustrated in
fig.~\ref{spower0}.
The $M(\vec{x}_0,R)$ versus $R$ figure is drawn as follows: For a
given $\vec{x}_0$ of a specific configuration we find $R$ of the
cluster in which it belongs, taken $\simeq \sqrt[3]{V}$, and we
calculate the integral $\int |\sigma(\vec{x})\sigma(\vec{x}_0)|\,
d^3x$, thus acquiring one point in the $M(\vec{x}_0,R)$ vs $R$
figure. For the same $\vec{x}_0$ we repeat this procedure until we
cover the whole ensemble, and the aforementioned figure is formed.
Averaging in $\vec{x}_0$ leads only to a slight modification of
the result, since $M(\vec{x}_0,R)\approx M(\vec{x}_0+\vec{l},R)$,
with $\vec{l}$ spanning the entire lattice, therefore in the
following we replace $M(\vec{x}_0,R)$ by $M(R)$. We observe that
in the log-log plot of $M(R)$ vs $R$, the slope $\Psi$, i.e the
fractal mass dimension $D_f$ according to (\ref{masspower}),
deviates from the theoretically expected value of $5/2$ by less
than 1\%. Furthermore, the mean value of the $\s$-field (spatial
average) is almost zero as expected to happen near the critical
point.

With this procedure we acquire an ensemble of field configurations
and the expected power law arises as a statistical property after
ensemble averaging \cite{manoscomput}. Alternatively, one could
extend the notion of the fractal dimension using individual
configurations, where the quantity of interest is the exponent
$d_f$ of the power law of the integral
\begin{equation}
m(R)=\int_{R} |\sigma(\vec{x})\sigma(\vec{x}_0)|\,
d^3x\,d^3x_0\sim R^{d_f}. \label{masspowereachconf}
\end{equation}
Calculating $d_f$ for each configuration we obtain a distribution
around $5/2$ with standard deviation $\approx0.05$.
 As expected the ensemble average of
$d_f$ is $\langle d_f\rangle=D_f$, within an error of less than
0.5\%.

\subsection{Generation of initial thermal $\pi$-configurations}

We generalize the method of \cite{Cooper} in order to produce an
ensemble of 3-D $\pi$-configurations in real space, corresponding
to an ideal gas at temperature $T_0$.
 The unperturbed Hamiltonian for the classical scalar field
theory in three dimensions is
\begin{equation}
H=\frac{1}{2}\int
d^3x[(\partial_t\pi(\vec{x},t))^2+(\nabla\pi(\vec{x},t))^2+m_\pi^2\pi(\vec{x},t)^2].
%-\frac{\la}{8}\pi(x,t)^4.
\label{ham}
\end{equation}

The free particle solutions for $t=0$ are
\begin{eqnarray}
\pi(\vec{x},0)=\int^{+\infty}_{-\infty}\frac{d^3k}{(2\pi)^3}\pi_{k0}\,e^{i\vec{k}\vec{x}}=
\int^{+\infty}_{-\infty}\frac{d^3k}{(2\pi)^3}\frac{(a_k+a_{-k}^\ast)}{\sqrt{2\omega_k}}e^{i\vec{k}\vec{x}}
\nonumber\\
\dot{\pi}(\vec{x},0)=\int^{+\infty}_{-\infty}\frac{d^3k}{(2\pi)^3}\xi_{k0}\,e^{i\vec{k}\vec{x}}=
\int^{+\infty}_{-\infty}\frac{d^3k}{(2\pi)^3}\sqrt{\frac{\omega_k}{2}}i(a_{-k}^\ast-a_k)e^{i\vec{k}\vec{x}}.
\label{sol}
\end{eqnarray}
where $\omega_k=\sqrt{k^2+m_\pi^2}$.

Now, choosing an initial classical density distribution
\cite{Cooper}
\begin{equation}
\rho[\pi,\dot{\pi}]=Z^{-1}(\beta_0)\,
\exp{\left\{-\beta_0\,H[\pi,\dot{\pi}]\right\}}, \nonumber
\end{equation}
 and substituting the Hamiltonian (\ref{ham})
with the free particle solutions (\ref{sol}), we finally get
\begin{equation}
\rho[x_k,y_k]=Z^{-1}(\beta_0)\,
\exp{\left\{-\beta_0\,\int^{+\infty}_{-\infty}\frac{d^3k}{(2\pi)^3}\omega_k(x_k^2+y_k^2)\right\}},
\label{findens}
\end{equation}
with $\beta_0=1/T_0$, and where: $a_k=x_k+iy_k$ with $x_k, y_k$
real. In order to produce a thermal ensemble (at temperature
$T_0$) of configurations for $\pi(\vec{x},0)$ and
$\dot{\pi}(\vec{x},0)$, we select $x_k$ and $y_k$ from the
gaussian distribution (\ref{findens}), assemble $a_k$ and then
substitute in (\ref{sol}). Lastly, since we have three components
of the pion pseudoscalar field, we independently repeat this
procedure accordingly. All the characteristics of the
$\pi$-ensemble such as the correlation function
$\langle\pi(\vec{x})\pi(\vec{x}+\delta
\vec{x})\rangle-\langle\pi(\vec{x})\rangle\langle\pi(\vec{x}+\delta
\vec{x})\rangle$, which turns out to be a $\delta$-function, are
consistent with the assumption of an ideal thermal gas.

\section{Numerical Solutions}

We study the evolution of the system determined by equations
(\ref{eom1}) which we solve in 3-D $25\times25\times25$ lattice,
using as initial conditions an ensemble of $10^4$ independent
$\s$-configurations on the lattice generated as described above,
i.e possessing fractal characteristics, and  $10^4$ configurations
for each $\pi$ component corresponding to an ideal gas at
temperature $T_0\approx140$ MeV. The initial time derivatives of
the $\s$-field, forming the kinetic energy, are assumed to be
zero, since this is a strong requirement of the initial
equilibrium. The used population is by far satisfactory since the
results converge for ensembles with more than $7\times10^3$
configurations (numerically tested) for the considered lattices
(sizes from $15\times15\times15$ to $25\times25\times25$ sites).
In addition, we find that the obtained results are independent of
the lattice spacing $\alpha$ despite the discontinuities in the
derivatives of the piecewise constant configurations, as the
corresponding variation $\delta \s$ goes to zero fast enough so
that the limit $\lim_{\alpha\rightarrow0}\frac{\delta\s}{\alpha}$
exists \cite{manoscomput}. Lastly, varying the value of $T_0$
between 100 MeV and 180 MeV it has a negligible effect on the
results. We investigate the evolution of  $M(R)$ of the whole
ensemble which initially follows a power law $\sim R^{\Psi(0)}$
with $\Psi(0)\equiv D_f=5/2$. We also study the evolution of
$m(R)$ for each configuration. At $t=0$ $m(R)$ possess a power-law
behavior of the form $\sim R^{\psi(0)}$, with the exponent
$\psi(0)\equiv d_f$ normally distributed around $5/2$ (with
standard deviation $\approx0.05$).

\begin{figure}[h]
\begin{center}
\mbox{\epsfig{figure=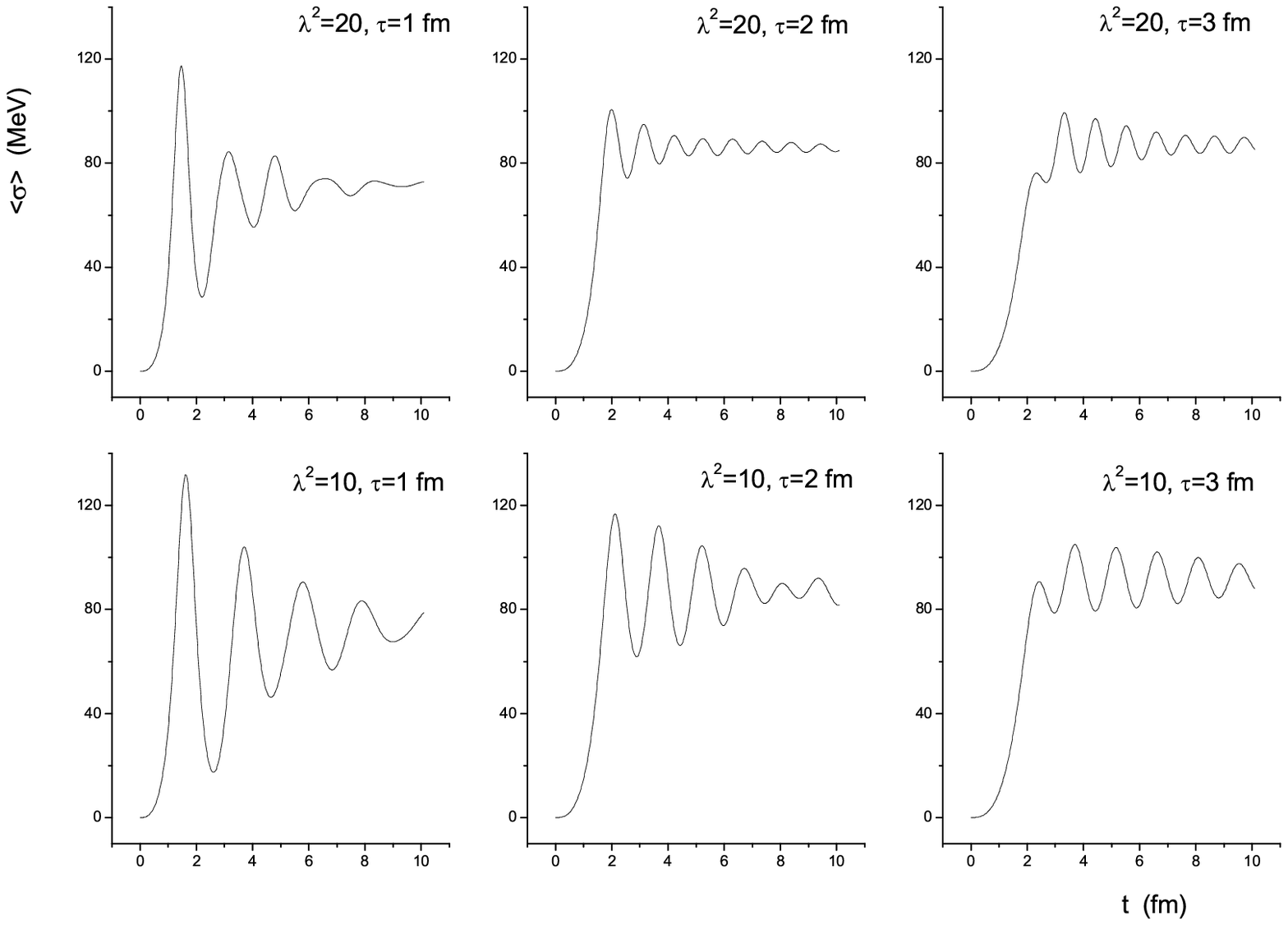,width=14.3cm,angle=0}}
\caption{\it Time evolution of the mean field value
$\langle\s\rangle$ for various $\la^2$ and $\tau$ cases.}
\label{sfield}
 \end{center}
 \end{figure}

\begin{figure}[h]
\begin{center}
\mbox{\epsfig{figure=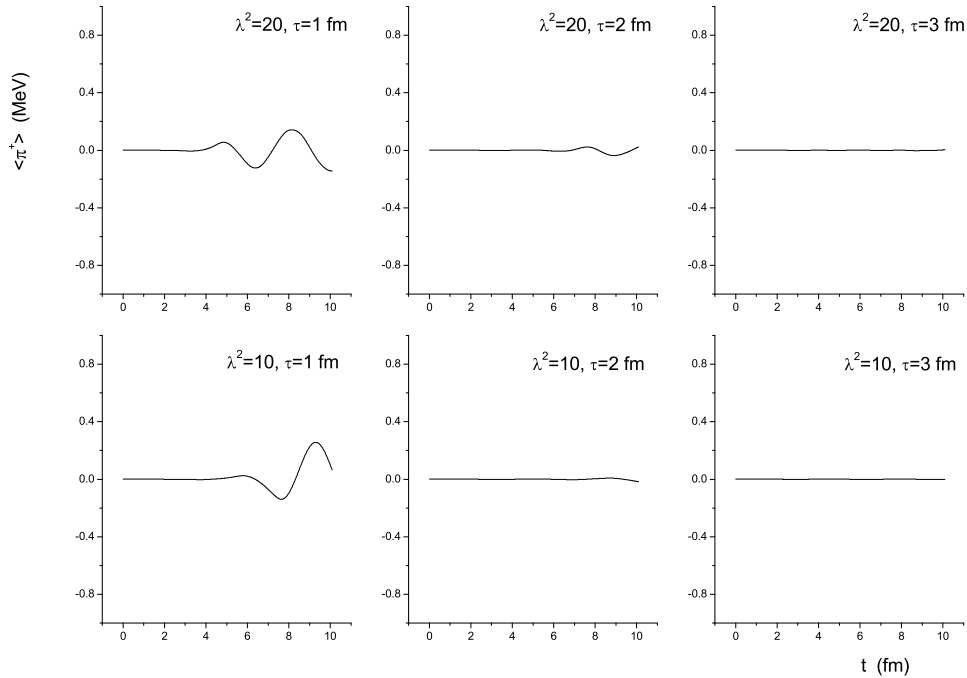,width=14.3cm,angle=0}}
\caption{\it Time evolution of the mean field value
$\langle\pi\rangle$ for various $\la^2$ and $\tau$ cases.}
\label{pfield}
 \end{center}
 \end{figure}

We evolve the system for various $\la^2$, corresponding to
different $m_\s$ values at $T=0$, and for various quench times
$\tau$. In figs.~\ref{sfield} and \ref{pfield} we demonstrate the
evolution of the mean field values $\langle\s\rangle$ and
$\langle\pi^+\rangle$ (the other components are similar), where
the averages are taken over all statistically independent
configurations.
As expected $\langle\s\rangle$ oscillates around the potential
minimum which relaxes after $\tau$, while $\langle\pi^+\rangle$
stays around zero due to the absence of a linear in $\vec{\pi}$
term in the potential (\ref{pot}).

\begin{figure}[!]
\begin{center}
\mbox{\epsfig{figure=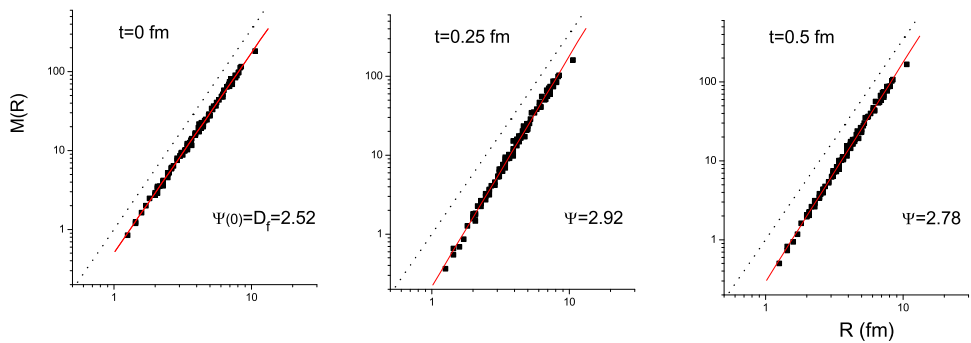,width=14.3cm,angle=0}}
\caption{\it $M(R)$ vs $R$ for three successive times, $t=0$ fm,
$t=0.25$ fm and $t=0.5$ fm, for typical $\la^2$ and $\tau$ values.
The solid lines mark the power-law fit, and the dotted line has
slope $2.5$ and is used as a reference.} \label{successlopes}
 \end{center}
 \end{figure}

\begin{figure}[!]
\begin{center}
\mbox{\epsfig{figure=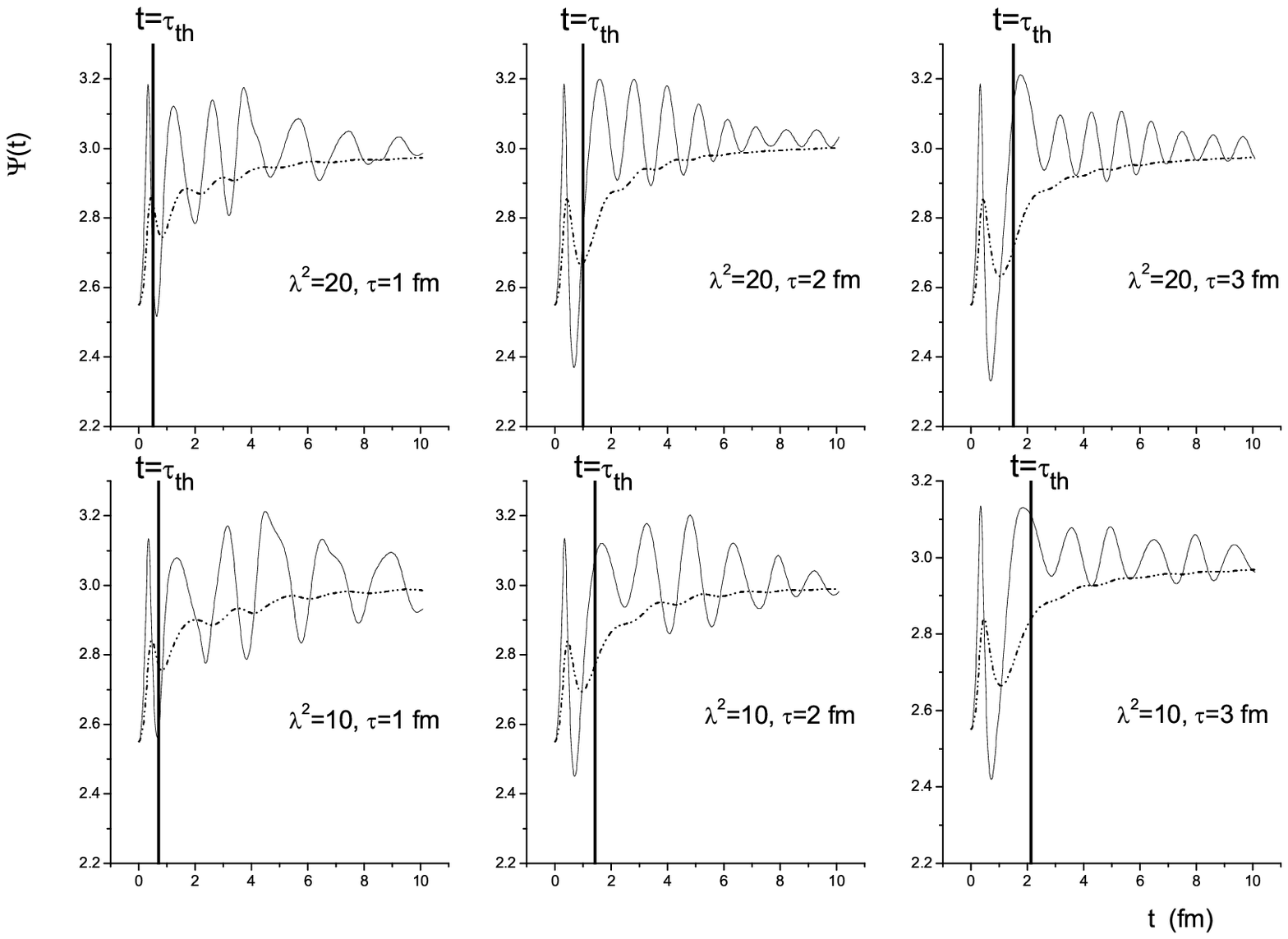,width=17cm,angle=0}}
\caption{\it Time evolution of $\Psi(t)$ (solid line) and
$\langle\Psi\rangle_t$ (dashed line) for the whole ensemble, for
the same $\la^2$ and $\tau$ cases as before. The vertical line
marks the time $t=\tau_{th}$ where $m_\s$ reaches the threshold
$2m_\pi$.} \label{slopetotal}
 \end{center}
 \end{figure}

In fig.~\ref{successlopes} we present $M(R)$ vs $R$ for three
successive times and we observe that the slope $\Psi(t)$
fluctuates, leading to a distortion of the initial fractal
geometry.
In fig.~\ref{slopetotal} we depict the evolution of the slope
$\Psi(t)$ of $M(R)$ versus $R$  (each $\Psi(t)$ value obtained
trough a linear fit), for the same $\la^2$ and $\tau$ values as
before.
With the dashed line we plot the time average of $\Psi(t)$ defined
by $\langle\Psi\rangle_t=\frac{1}{t}\int_0^t\Psi(t')dt'$. We
observe the remarkable phenomenon that the characteristic exponent
$\Psi(t)$ after reaching the value of the embedding dimension 3,
it fluctuates and for particular times it becomes almost equal to
$\Psi(0)=5/2$. Thus, after the first deformation, the initial
critical behavior of the whole ensemble is partially restored and
deformed repeatedly. A detailed explanation of this revival is
given in \cite{deterministic}. The key point is that the partial
restoration takes place when $\langle\s(t)\rangle$ passes through
its lower turning point, where the $\s$-field (seen as a system of
coupled oscillators) reaches a state similar to the initial one.
This phenomenon weakens gradually, and finally the dynamics
dilutes completely the initial critical behavior.

\begin{figure}[h]
\begin{center}
\mbox{\epsfig{figure=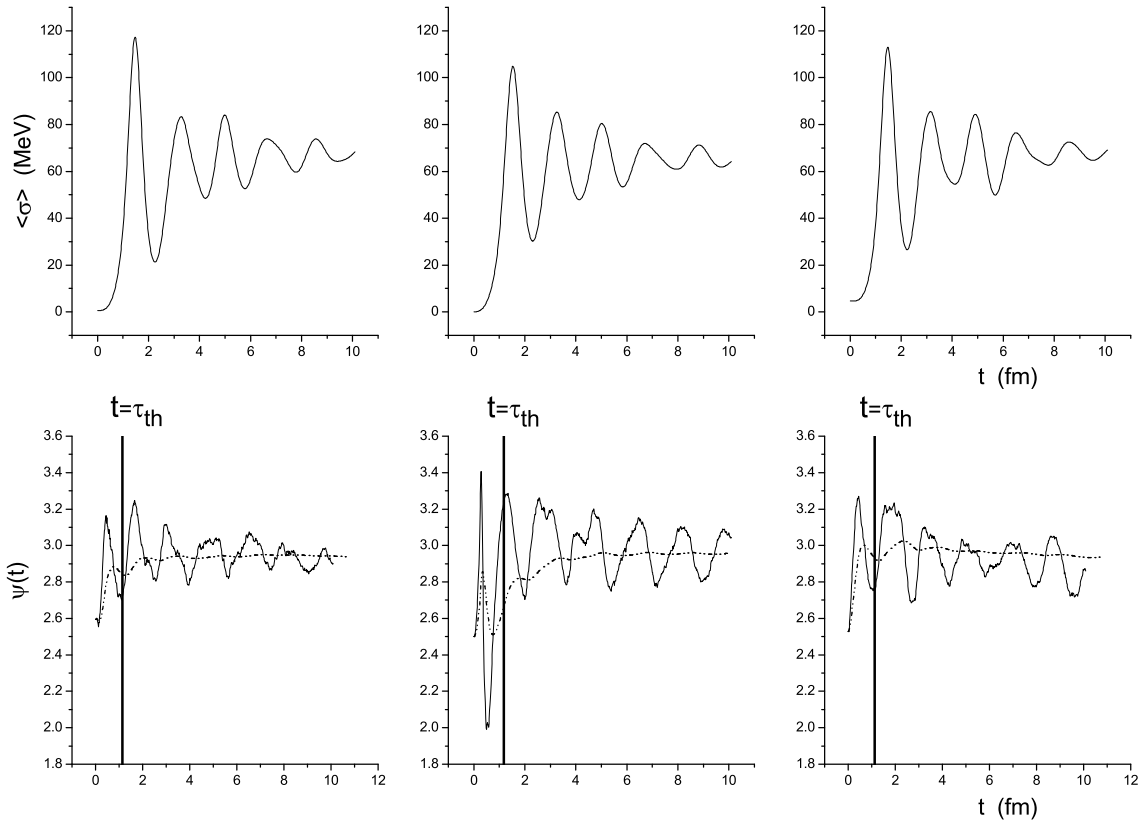,width=17cm,angle=0}}
\caption{\it $\langle\s\rangle$, $\psi(t)$ (solid line) and
$\langle\psi\rangle_t$ (dashed line) for three independent
configurations, for $\la^2=20$ and $\tau=1$ fm. The vertical line
marks the time $t=\tau_{th}$ where $m_\s$ reaches the threshold
$2m_\pi$.} \label{1confslope}
 \end{center}
 \end{figure}

This phenomenon is also visible in the evolution of the slope
$\psi(t)$ of $m(R)$ versus $R$, for each configuration. Indeed, in
fig.~\ref{1confslope} we demonstrate the evolution of the mean
field value as well as of $\psi$ and of its time average, for
$\la^2=20$ and $\tau=1$ fm, for three independent configurations
of the ensemble, corresponding to initial value  $\psi(0)=d_f$
equal to 2.51, 2.50 and 2.53 respectively.
We observe an oscillatory behavior of $\psi(t)$ similar to that of
$\Psi(t)$ of the whole ensemble presented in
fig.~\ref{slopetotal}.

In a heavy ion collision, if the fireball passes near the critical
point we expect $\s$ to have almost zero mass. As the system
expands, its temperature decreases, $v$ increases towards its zero
temperature value $v_0$ and $m_\s=\sqrt{2\lambda^2 v(t)^2}$ tends
to its freeze out value. However, when it reaches the threshold
$m_\s=2m_\pi$, at time $t=\tau_{th}$, it starts decaying into
pions, and the decay rate of $\s\rightarrow\pi^{+}\pi^{-}$, being
proportional to the available phase space parameter
$\sqrt{1-\frac{4 m_{\pi}^2}{m_\s^2}}$, becomes larger as $m_\s$
increases. So if the critical characteristics of $\s$ have
survived at that threshold, they can be transferred to the
produced pions, leaving signatures of the critical point at the
detectors. Note here that these pions are not affected by the
thermal ones which constitute the environment. In
figs.~\ref{slopetotal} and \ref{1confslope} the vertical line
depicts the threshold time $t=\tau_{th}$. We are interested in
investigating the time averaged measures after this threshold
since these quantities are observable.

The first measure one has to look at is the evolution of $\Psi(t)$
and $\langle\Psi\rangle_t$ for the whole ensemble, after
$t=\tau_{th}$, as it is shown in fig.~\ref{slopetotal}. As we
observe these vary between 2.6 and 2.9, for the various $\la^2$
and $\tau$ values, offering weak traces of the initial power law
of 5/2. This cogitation is amplified by the fact that if we evolve
our system with conventional initial conditions ($D_f\approx3$),
then $\Psi(t)$ and $\langle\Psi\rangle_t$ remain always
$\approx3$, as we have tested, supporting the assumptions that
slopes $<2.9$ indicate initial critical behavior. However, slopes
close to 3 could originate from conventional strong processes
leading to power-law correlations with very small intermittency
exponents \cite{Bialas86}. Therefore we have to refer to more
sophisticated measures in order to acquire a direct observation of
the QCD isothermal critical exponent $\delta$, independently of
the specific model. One way to achieve this is to perform
event-by-event analysis of our results.

\begin{figure}[!]
\begin{center}
\mbox{\epsfig{figure=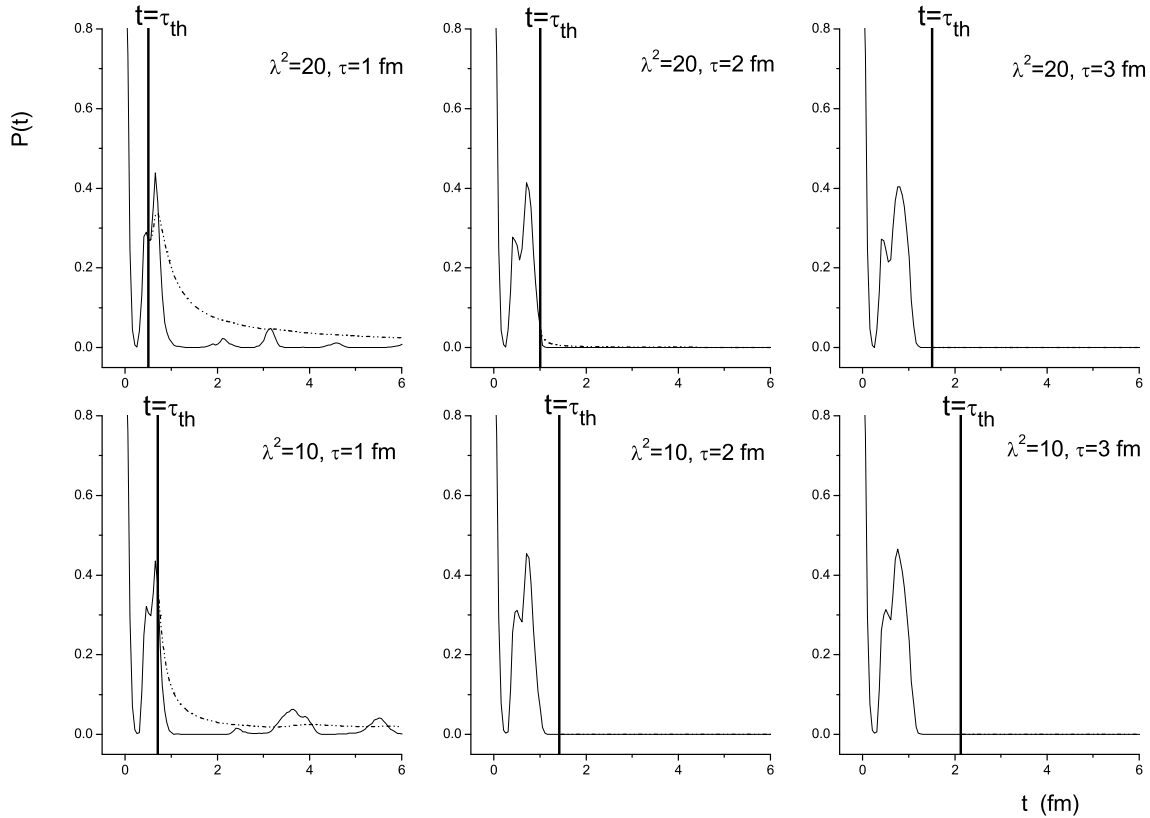,width=17cm,angle=0}}
\caption{\it Time evolution of the percentage $P(t)$ of the
configurations that reacquire the value
$\psi(t)\approx\psi(0)=d_f\approx5/2$ at $t$. With the dashed line
we depict its time average $\langle P\rangle_t$ after
$t=\tau_{th}$.} \label{eventbyevent}
 \end{center}
 \end{figure}

As we have mentioned, we use an ensemble of $\s$-field
configurations, each one possessing its own $d_f$ which can vary,
leading to a distribution with mean value $\approx5/2$ and
standard deviation $\approx0.05$. An event-by-event analysis of
the system evolution consists in calculating the percentage $P$ of
the initial configurations (all of which have
$\psi(0)=d_f\approx5/2$) that possess again this value at time $t$
(actually we count those with $2.4<\psi(t)<2.6$). Obviously,
$P(t)$ leads to experimentally accessible effects after
$t=\tau_{th}$. In fig.~\ref{eventbyevent} we depict the evolution
of $P(t)$, and its time average $\langle P\rangle_t$ for
$t>\tau_{th}$ (the vertical line marks $t=\tau_{th}$), for the
same $\la^2$ and $\tau$ values described above.
As we observe, $P(t)$ is quite large and decreases with time as
expected, while it presents some fluctuations which correspond to
the peaks of $\Psi(t)$ evolution of fig.~\ref{slopetotal}. As we
mentioned above we can define the relaxation time-scale
$\tau_{rel}$, after which the initial critical behavior is
completely lost, as the time where $\langle P\rangle_t$ becomes
less than 0.5\%. Therefore, if $\tau_{rel}$ is larger than
$\tau_{th}$ then the critical characteristics will be transferred
to the produced, through the decay $\s\rightarrow\pi^{+}\pi^{-}$,
pions.

\begin{figure}[!]
\begin{center}
\mbox{\epsfig{figure=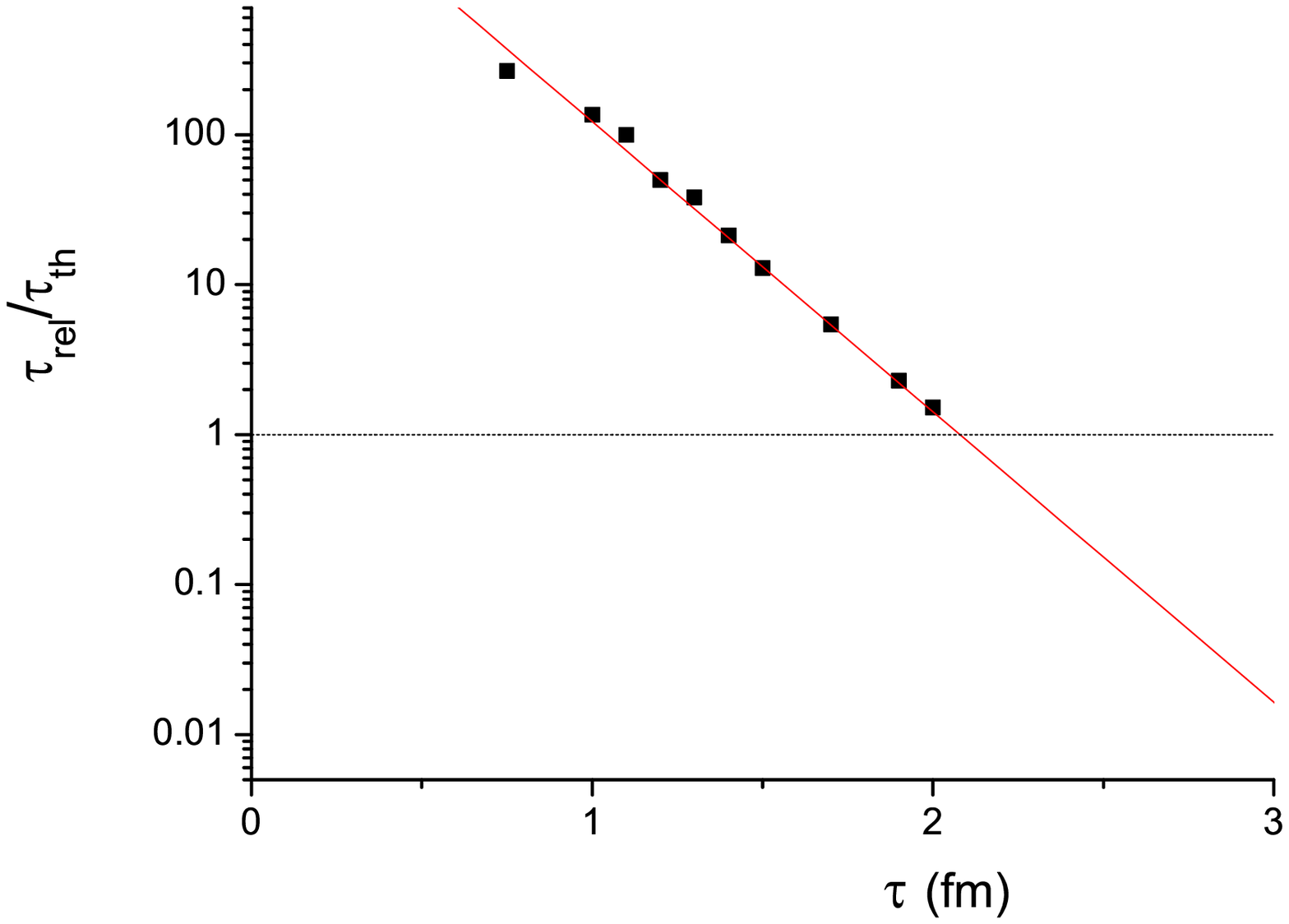,width=11cm,angle=0}}
\caption{\it The dependence of the ratio $\tau_{rel}/\tau_{th}$ on
the quench duration $\tau$ for $\lambda^2=20$. The dashed line
depicts an exponential fit.} \label{treltthla20}
 \end{center}
 \end{figure}

The quench duration seems to affect the results significantly. For
quench time $\tau=1$ fm the aforementioned time averaged $\langle
P\rangle_t$ can be quite large for the first 1-2 fm after
$t=\tau_{th}$, as can be observed in fig.~\ref{eventbyevent}, and
it reaches the cut 0.5\% at $\tau_{rel}\approx40$ fm, a clearly
larger value than the expected freeze-out time in heavy ion
collisions. In other words, $\tau_{rel}\approx40$ fm means that
$\langle P\rangle_t>0.5$\% at the freezeout. For $\tau=2$ fm,
$\langle P\rangle_t$ has a similar behavior but $\tau_{rel}$
decreases to $\approx1.6$ fm, for the $\lambda^2=20$ case. For
$\tau=3$ fm, $\langle P\rangle_t$ is always below 0.5\% since in
this case $t=\tau_{th}$ is reached after the first large
fluctuation of $\Psi(t)$, as it is shown in the e and f plots of
fig.~\ref{slopetotal}. Thus, in this case, according to the above
definition $\tau_{rel}=0$. In general for faster quenches,
$\langle P\rangle_t$, i.e the time average of the percentage of
the field configurations that reacquire the value
$\psi(t)\approx5/2$ after $t=\tau_{th}$, is significantly larger.
On the other hand, for slower quenches $\tau_{th}$ increases and
it supplants $\tau_{rel}$, thus the dynamics dilute the traces of
the initial critical profile, before the decay
$\s\rightarrow\pi^{+}\pi^{-}$ activates.
 The dependence of the ratio $\tau_{rel}/\tau_{th}$ on the quench
duration $\tau$ is depicted in fig.~\ref{treltthla20} for the
$\lambda^2=20$ case.
As we observe, for a wide range of $\tau$ values
($\tau\lesssim2.1$ fm in the $\lambda^2=20$ case),
$\tau_{rel}/\tau_{th}>1$ and especially for small $\tau$ it can be
sufficiently large ($\tau_{rel}/\tau_{th}\simeq10$ for
$\tau\simeq1.6$ fm). As the model parameter $\tau$ increases, the
ratio $\tau_{rel}/\tau_{th}$ decreases almost exponentially as can
be induced from the exponential fit presented with the dashed line
in fig.~\ref{treltthla20}.

The effect of $\lambda^2$ (that is the value of the $m_\s$ at
$T=0$) is not so crucial and $\tau_{rel}$ shows a small decrease
with decreasing $\lambda^2$, while $\tau_{th}$ increases slightly.
However, for a given value of the interval $\tau_{rel}-\tau_{th}$
the pion production from the critical $\s$'s is controlled by
$\lambda^2$. Since the $\s$-decay rate is larger for larger
$m_\s/m_\pi$, we expect faster decay for $\la^2=20$ compared to
$\la^2=10$, since in the former case the final $m_\s$ is larger.
Experimentally we expect a larger number of produced pions in the
$\la^2=20$ case. Inversely, when $\la^2=20$ a smaller interval
$\tau_{rel}-\tau_{th}$ is needed in order to produce a given
number of pions through the decay of critical $\s$'s. This is the
main difference of the two $\la^2$ cases, although the
corresponding plots look alike.

The discussion above reveals that the critical behavior of the
$\s$-field will sustain, in a significant percentage of the
initial configurations, for times after the $\s$'s start to decay,
thus transferring the critical profile to the produced pions,
despite the abrupt non-equilibrium evolution from its initial
equilibrated state.

It is important to exclude the inverse possibility, i.e appearance
of events with fractal geometry being generated out of the
evolution of conventional initial conditions. Therefore, we
perform the following test: We evolve our system using an ensemble
of $\s$-configurations with random initial conditions. This
conventional profile always possesses $\langle d_f\rangle\approx3$
with a very small standard deviation ($\approx0.02$). In this case
not even one configuration (out of $10^4$) acquires slope
$\lesssim2.96$ ever, while the distribution becomes even narrower
as times passes. As a consequence, $P(t)$ and $\langle P\rangle_t$
are always exactly equal to zero, within the used population.
Therefore, the detection of events with critical profile at the
freeze out, will offer safe signatures of the initial criticality
of the system. The requirement $\langle P\rangle_t\geq0.5\%$
remains robust for a long time and for a variety of $\la^2$ and
$\tau$ as we have shown, providing a guarantee that such events
are very likely to be observed in experiments of relatively high
statistics.

The above treatment shows, together with other approaches in the
past \cite{Antoniou2001,hatta03,SRS2,Randrup00,biro97}, that a
systematic study of event-by-event fluctuations in relativistic
nuclear collisions gives rise to a powerful tool in the search for
novel, unconventional effects associated with QCD phase
transitions (critical fluctuations, disoriented chiral
condensates, dissipation at the chiral phase transition). In this
work we have examined the evolution of critical fluctuations
neglecting the effects of dissipation in this out-of-equilibrium
process. Although a detailed study of dissipative aspects of this
evolution is beyond the scope of this work, an estimate of their
influence on the fractal structure of critical fluctuations is
necessary before drawing our final conclusions. For this purpose
we have considered a simplified modification of $\s$ field
dynamics adding a dissipative term
$\Gamma\dot{\s}(\vec{x},t)-\gamma\nu(t)$ in the first equation
(\ref{eom1}). $\nu(t)$ is the noise term fulfilling the relations
$\langle\nu(t)\rangle=0$ and
$\langle\nu(t)\nu(t')\rangle=\gamma^2\delta(t-t')$.
 The coefficients
$\Gamma$ and $\gamma$ are related through the formula
$\gamma=\sqrt{2\Gamma T/V}$, resulting from the
fluctuation-dissipation theorem, with $T$ the environment
temperature and $V$ the system volume. The friction coefficient
$\Gamma$ introduces a new time scale in the process
($\tau_{f}=\Gamma^{-1}$) associated with the rolling down of the
effective potential \cite{biro97}. With this modification we have
solved the equations of motion (\ref{eom1}) following the method
discussed in the beginning of this section.

\begin{figure}[!]
\begin{center}
\mbox{\epsfig{figure=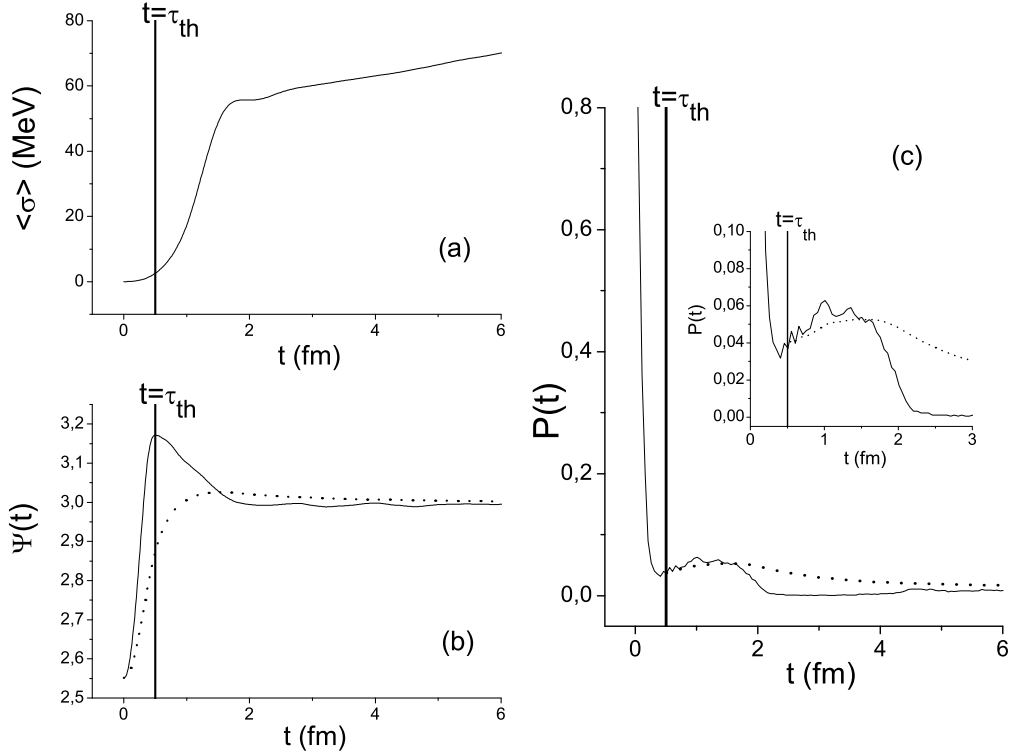,width=15cm,angle=0}}
\caption{\it $\langle\s\rangle$, $\psi(t)$ (solid line),
$\langle\psi\rangle_t$ (dotted line) and $P(t)$ (solid line),
$\langle P\rangle_t$ (dotted line), in the strong dissipative case
($\Gamma^{-1}\simeq0.5$ fm), for $\la^2=20$ and $\tau=1$ fm.}
\label{posdiss}
\end{center}
\end{figure}

We use the typical values $\la^2=20$ and $\tau=1$ fm, and for the
friction we impose $\Gamma^{-1}\simeq0.5$ fm, which corresponds to
the strong friction regime \cite{biro97}. As we observe in
fig.~\ref{posdiss}, $\langle\s(t)\rangle$ and $\Psi(t)$ change
significantly compared to the non-dissipative case, and the traces
of the initial fractal geometry dilute faster.
However, the change in $\langle P\rangle_t$ does not spoil its
observability since $\tau_{th}$ and $\tau_{rel}$ are almost
unaffected (see the inset in fig.~\ref{posdiss}c). Hence,
dissipation influences the mean values evolution, but it does not
affect qualitatively the fluctuation-related measures such as
$\langle P\rangle_t$. Therefore, the signals of the initial
critical characteristics are still visible in the event-by-event
analysis. Moreover, we have found, as expected, that weakening the
dissipation, i.e decreasing $\Gamma$, its effects are suppressed
and the system evolution tends smoothly to that of the
non-dissipative case.

\section{Summary and Conclusions}

In this work we have explored the dynamics of critical
fluctuations which are expected to occur in the $\s$-mode, near
the QCD critical point. For this purpose we have adapted the
$\s$-model Lagrangian in order to describe correctly the
characteristics of the order parameter associated with the
critical endpoint of the QCD phase transition \cite{RW}. The issue
is of primary importance in the search for the existence and
location of the QCD critical point, in experiments with nuclei. At
the phenomenological level these fluctuations are expressed
through the fractal mass dimension of the $\s$-field
configurations, determining the properties of the condensate at
criticality \cite{Antoniou2001}. We have studied the evolution of
the initial critical characteristics of the $\s$-field in thermal
equilibrium and the possibility to reveal signals of critical
fluctuations not affected by the dynamics which drives the system
out-of-equilibrium. We found that for a wide range of quench
time-scales ($\tau$) and coupling values ($\lambda^2$), the
initial critical profiles may survive for a long time after the
system reaches the two-pion threshold value of the sigma mass,
$m_\s=2m_\pi$. This result is more transparent in an
event-by-event study of the phenomenon, where the evolution of
individual configurations is investigated.

As a consequence of this study, the fractal dimension of critical
fluctuations in QCD matter, turns out to be a remarkable index for
the location of the QCD critical point, in experiments with
nuclei, of relatively high statistics. In fact, it remains robust,
in a class of events, against dynamical effects and it leads to a
characteristic pattern of intermittent fluctuations
\cite{Antoniou2001,Bialas86} in transverse momentum space,
providing us with a signature of the critical point without
ambiguities owing to dynamics.\\

\paragraph*{{\bf{Acknowledgements:}}} We thank N. Tetradis for useful discussions. One of us (E.N.S) wishes to
thank the Greek State Scholarship's Foundation (IKY) for financial
support. The authors acknowledge partial financial support through
the research programs ``Pythagoras'' of the EPEAEK II (European
Union and the Greek Ministry of Education) and ``Kapodistrias'' of
the University of Athens.

\end{document}